\documentclass[prb,twocolumn,showpacs,amsmath,amssymb,superscriptaddress]{revtex4-1}
\usepackage{amsmath,amssymb,amsfonts,bm,color,graphicx,tabularx,physics}
\usepackage[unicode=true,colorlinks=true]{hyperref}
\usepackage[etex=true,export]{adjustbox}
\hypersetup{linkcolor=blue,citecolor=blue,urlcolor=blue}
\usepackage{tabularx,multirow,array,diagbox}
\usepackage{adjustbox}

\newcommand{\beq}{\begin{equation}}
\newcommand{\eeq}{\end{equation}}
\newcommand{\beql}{\begin{equation*}}
\newcommand{\eeql}{\end{equation*}}
\newcommand{\beqn}{\begin{eqnarray}}
\newcommand{\eeqn}{\end{eqnarray}}

\newcommand{\sgn}{{\rm sgn}}

\begin{document}

\title{ Higher-order topological superconductors based on weak topological insulators}
\author{Xun-Jiang Luo}
\author{Xiao-Hong Pan}
\affiliation{School of Physics, Huazhong University of Science and Technology, Wuhan, Hubei 430074, China}
\author{Xin Liu}
\email{phyliuxin@hust.edu.cn}
\affiliation{School of Physics, Huazhong University of Science and Technology, Wuhan, Hubei 430074, China}
\affiliation{Wuhan National High Magnetic Field Center and Hubei Key Laboratory of Gravitation and Quantum Physics, Wuhan, Hubei 430074, China}

\begin{abstract}
High-order topological phases host robust boundary states at the boundary of the boundary, which can be interpreted from their boundary topology. In this work, considering the interplay between superconductors and magnetic fields to gap the surface states of three-dimensional weak topological insulators, we show that second-order topological superconductors (TSCs) featuring helical or chiral Majorana hinge modes and third-order TSC featuring  Majorana corner  modes can be realized. Remarkably, the higher-order TSCs in our models can be attributed to their certain boundaries, surfaces or hinges, which naturally behave as first-order TSC in DIII or D symmetry class. Correspondingly, these higher-order TSCs can be characterized by the boundary first-order topological invariants, such as surface Chern numbers or surface $Z_2$ topological invariants for surface TSCs. Our models can effectively capture the topology of iron-based superconductors with desired inverted band structures and superconducting pairings.
\end{abstract}
\maketitle

\section{Introduction}
Topological insulators (TIs) and TSCs feature gapless boundary states protected by bulk topological property, which can be characterized by topological invariants defined throughout the Brillouin zone\cite{Qi2011,Hasan2010}. For concreteness, three-dimensional (3D) topological insulators (TIs) with time-reversal symmetry are known to be characterized by one strong and three weak $Z_2$ topological indexes \cite{Fu2007,Fu2007a}. According to these four $Z_2$ indexes, two classes TIs can be distinguished, strong TIs and weak TIs, which host an odd and even number of Dirac cones on surfaces, respectively. Remarkably, the findings of strong TIs\cite{Zhang2009, Chen2009,Hsieh2009} provide the parent materials of laboratory achievable TSCs. Several strong TI-based experimental platforms to realize TSCs have been proposed by taking advantage of the spin-momentum locking property of surface states of strong TIs\cite{Fu2008,Liu2011,Xu2014,Wang2015,He2017}.
To be specific, Majorana zero modes can be realized by performing the Fu-Kane scheme\cite{Fu2008}, which considers the $\pi$ flux vortex of the surface states with s-wave pairing. Moreover, chiral TSCs hosting chiral Majorana modes can be realized by considering magnetic TI thin film proximity to s-wave superconductor\cite{Wang2015, He2017}. The interplay between magnetic fields and superconductors to gap the surface states of strong TIs tremendously enriches the field of TSCs for their creation, detection, and manipulation \cite{Fu2009,Fu2009a,Akhmerov2009,Mi2013,Beenakker2019}. Meanwhile, this also naturally raises a question, can weak TIs, hosting more abundant surface states, be utilized to realize TSC?



On the other hand, the concept of topological phases has been generalized to higher-order over the past few years, which has been drawing great research interest as new topological phases of matter \cite{Benalcazar2017,Benalcazar2017a,Song2017,Langbehn2017,Geier2018,Khalaf2018,Zhu2019,Trifunovic2019,Schindler2018,
Geier2018,Schindler2018a,Ezawa2018,
Sheng2019,Wang2020a}. Generally, an nth-order topology in the dD system manifests itself by localized states at its (d-n)D boundaries. For instance, a second-order and third-order topological phases of a 3D system feature hinge modes and corner modes, respectively. In this classification scheme,
conventional topological phases are the first-order in nature. A prototypical method of obtaining higher-order topological phases is to gap adjacent gapless boundary
states in a nontrivial way, giving rise to boundary topological defects\cite{Song2017,Langbehn2017,Geier2018,Khalaf2018,Wang2020a,Zhu2018,
Volpez2019,Chen2020,Ren2020,Xu2019,Yue2019,Zhang2020b,
Yan2018,Wang2018,Zhang2019,Chen2019,Wu2020,Pan2019,Wu2019,Zhang2019a,Peng2019,Chen2019,Wu2020a}. This scenario can be realized by considering the gapless boundary states of TIs in 2D or 3D, which can be gapped by Zeeman fields breaking time-reversal symmetry, or superconductors breaking $\text{U}(1)$ symmetry. Correspondingly, the boundary domain walls, which can be generated by opposite magnetic gaps\cite{Zhu2018,Volpez2019,Chen2020,Ren2020,Xu2019,Yue2019,Zhang2020b}, superconducting gaps\cite{Yan2018,Wang2018,Zhang2019,Wu2020,Chen2019}, or magnetic-superconducting gaps\cite{Chen2019,Pan2019,Wu2019,Zhang2019a,Peng2019,Wu2020a} along adjacent boundaries, will lead to corner or hinge modes.
An alternative method to realize higher-order topological phases is to perform the boundary first-order topological phases\cite{Benalcazar2017,Benalcazar2017a,Shapourian2018,Chen2019a,Li2020}. Such as the well-known Benalcazar-Bernevig-Hughes model\cite{Benalcazar2017,Benalcazar2017a}, the edges of it behave as a 1D TI\cite{Benalcazar2017a,Li2020a}.

In this work, we start from weak TIs rather than strong TIs to design second-order and third-order TSCs hosting robust Majorana hinge modes and Majorana corner modes (MCMs), respectively. Concretely, we consider weak TIs with band inversion at both $\Gamma(0,0,0)$ and $\text{Z}(0,0,\pi)$ points in Brillouin zone. For this phase,
the surface Dirac cones only appear on side faces and their number is two.
We find that second-order TSC hosting helical Majorana hinge modes (HMHMs) can be realized in this system by considering $s_{\pm}$-wave superconductivity to gap the Dirac cones on the side faces.
Through the topological analysis of surfaces, we show that the corresponding side faces of this system naturally behave as a time-reversal invariant TSC (TRITSC), characterized by a surface $Z_2$ topological invariant. Moreover, in the presence of an in-plane Zeeman field, the side faces can be selectively driven to a chiral TSC through a surface topological phase transition, giving rise to chiral Majorana hinge modes (CMHMs), which is characterized by nontrivial surface Chern number. Analogously, third-order TSC can be realized  by designing
 1D hinge TSC. We propose a theoretical model to realize third-order TSC through this visualized principle and provide topological characterizations. For candidate material, our model can capture the topological property of iron-based superconductor Li(OH)FeSe, whose normal state was predicted to be a weak TI with band inversion at $\Gamma$ and $\text{Z}$ points\cite{Qin2019}.

\section{Second-order TSCs}
\label{M}
\subsection{Model and Hamiltonian}
We first introduce the Bogoliubov–de Gennes(BdG) Hamiltonian on a cubic lattice, which describes 3D weak TI with $s_{\pm}$-wave superconductor pairing,
\beqn
H_{\rm BdG}(\bm k)=\begin{pmatrix}
\label{TI}
\mathcal{H}( \bm{k})-\mu & -i\Delta(\bm{k})s_y\\
i\Delta(\bm{k})s_y&-\mathcal{H}^{\ast}(- \bm{k})+\mu\end{pmatrix},
\eeqn
with the normal-state Hamiltonian $\mathcal{H}( \bm{k})=\sum_{j=x,y,z}(m_0-t_j\cos k_j)\sigma_zs_0+\sin k_j\sigma_xs_j$. Here, $\mu$ is the chemical potential, $m_0$ and $t_j$ are taken to be positive model parameters, $s$ and $\sigma$ are Pauli matrices in the spin and orbital spaces, respectively. The $s_{\pm}$-wave pairing order parameter is $\Delta(\bm{k})=\Delta_0+\Delta_1(\cos k_x+\cos k_y)$, which can be achieved intrinsically in an iron-based superconductors\cite{Wu2016,Xu2016,Zhang2018,Wang2018b}.

\begin{figure}
\centering
\includegraphics[width=3.3in]{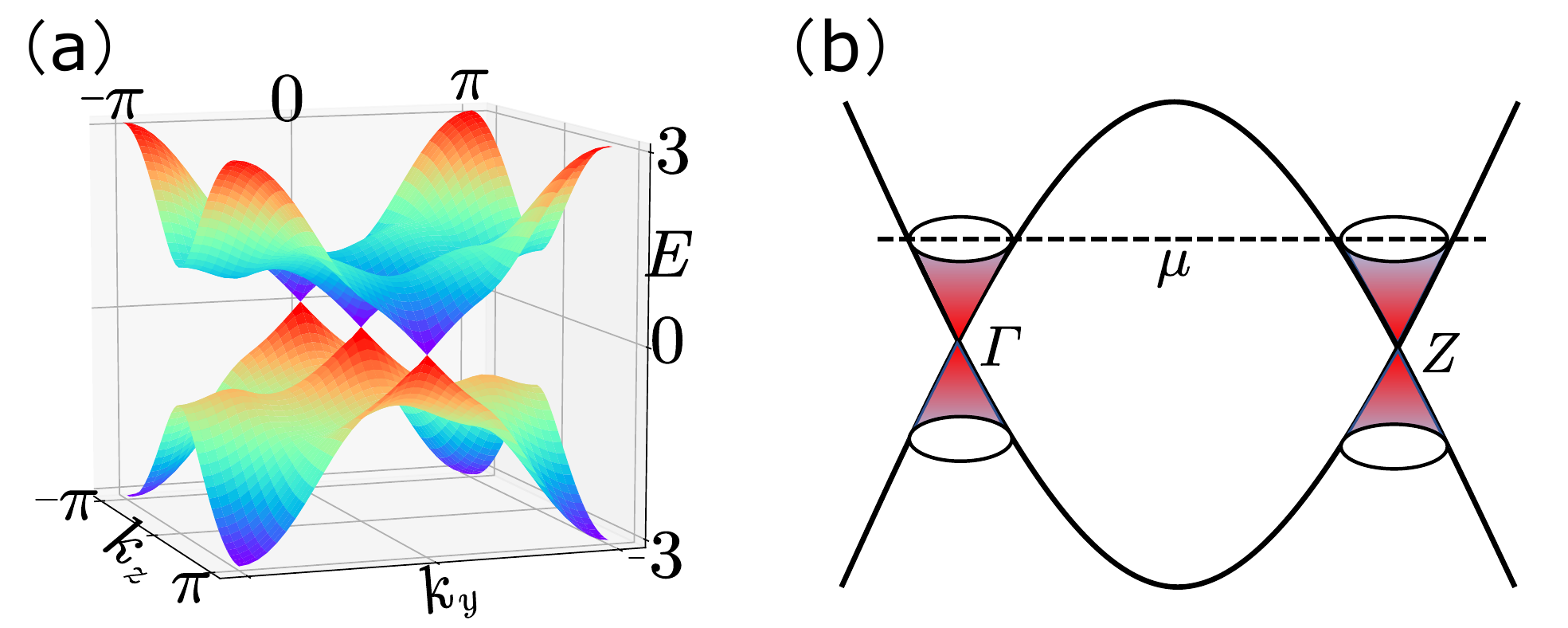}
\caption{(a): The energy spectrum of two Dirac cones on (100) surface is plotted. Model parameters are taken as $m=2,t_x=t_y=2,t_z=1$. (b): Schematic plot of the two Fermi surfaces around the two surface Dirac cones.}
\label{dirac}
\end{figure}

It is readily verified that $H_{\rm BdG}(\bm k)$ respects time-reversal symmetry ($\mathcal{T}=is_y\mathcal{K}$), inversion symmetry ($I=\sigma_z$), and particle-hole symmetry ($\mathcal{P}=\tau_x\mathcal{K}$), where $\tau$ are Pauli matrices in particle-hole space and $\mathcal{K}$ denotes the complex conjugation. Thus, the band topology of $\mathcal{H}( \bm{k})$ is easily read off using the Fu-Kane criterion\cite{Fu2007a} based on the eigenvalues of $I$ at the eight TRI momenta points. For our purposes, we focus on the region of the parameters, in which the band inversion of $\mathcal{H}( \bm{k})$ occurs at both $\Gamma$ and $\text{Z}$ points in the first Brillouin zone. Constrained into this region, there are no topological surface states on (001) surface while there are two Dirac cones on each side face, crossing at $k_z=0$ and $k_z=\pi$, respectively. For a slab geometry with open boundary condition along $x$ direction, the energy spectrum of surface Dirac cones on $(100)$ surface is shown in Fig.~\ref{dirac}(a).

Based on this weak TI, the physical picture of realizing surface TSC can be described as follows.
When the chemical potential is inside the bulk electronic bandgap, there are two Fermi surfaces around each Dirac cone, as shown in Fig.~\ref{dirac}(b). Further considering bulk $s_{\pm}$-wave pairing, surface Dirac cones of weak TI, described by 2D massless Dirac fermion, are gapped out by the superconducting pairing, which introduces Dirac
mass. Remarkably, different from the case of uniform s-wave pairing, $s_{\pm}$-pairing can gap the Dirac cones with different magnitudes. Owing to this property, the Dirac masses of the two Dirac cones on the same side face can have a sign change under appropriate parameters condition. In this case, the side face exists an odd number of Fermi surfaces with negative superconducting pairing, which enables the realization of a surface TRITSC\cite{Qi2010,Zhang2013,Haim2018}, although the bulk is a trivial superconductor. To demonstrate this simple physical picture, we perform surface theory in the following.

\subsection{Second-order TSC in DIII symmetry class}
For the sake of simplicity, we take $\mu=0$ and focus on the continuum model by expanding the bulk Hamiltonian $H_{\rm{BdG}}(\bm{k})$ to second-order around $\Gamma$ and $Z$ points, respectively
\beqn
&H_{\rm BdG}^{\gamma(z)}=(m_{\gamma(z)}+t_x/2k_x^2+t_y/2k_y^2\pm t_z/2k_z^2 )\tau_z\sigma_z\nonumber+k_x\sigma_xs_x\\
&+k_y\tau_z\sigma_xs_y\pm k_z\sigma_xs_z+[\bar{\Delta}-\Delta_1/2(k_x^2+k_y^2)]\tau_ys_y,
\eeqn
where bulk electronic bandgap $m_{\gamma(z)}=m_0-t_x-t_y\mp t_z<0$ are taken to ensure that the band inversion of normal states occurs at $\Gamma$ and $\text{Z}$ points simultaneously, $\bar{\Delta}$ is defined as $\Delta_0+2\Delta_1$. Without loss of generality, we concentrate on the topology of (100) surface and take the open boundary condition of $x$ direction. Under this condition,
we replace $k_x\rightarrow -i\partial_x$ and then the BdG Hamiltonian can be decomposed as $H_{\rm BdG}^{\gamma(z)}=H_0^{\gamma(z)}(x)+H_p^{\gamma(z)}(x,k_y,k_z)$, in which
\beqn
H_0^{\gamma(z)}=(m_{\gamma(z)}-\frac{t_x}{2}\partial_x^2)\tau_{z}\sigma_z-i\partial_x\sigma_x s_x,\nonumber\\
H_p^{\gamma(z)}=k_y\tau_z\sigma_x s_y\pm k_z\sigma_x s_z+\Delta(x)\tau_y s_y,
\eeqn
where $\Delta(x)$ is defined as $\bar{\Delta}+\frac{\Delta_1\partial_x^2}{2}$ and the irrelevant terms $k_y^2$ and $k_z^2$ have been omitted. Now we solve the zero energy states of $H_0^{\gamma(z)}$ and consider $H_p^{\gamma(z)}$ as a perturbation to extract the surface low energy Hamiltonian.

Solving the eigen equation $H_0^{\gamma(z)}|\psi_{\alpha}^{\gamma(z)}(x)\rangle=0$ under the boundary conditions $|\psi_{\alpha}^{\gamma(z)}(x=0)\rangle=|\psi_{\alpha}^{\gamma(z)}(x=-\infty)\rangle=0$, there are four zero energy states. The wave function of these zero energy states can be written as
\beqn
|\psi_{\alpha}^{\gamma(z)}(x)\rangle=\mathcal{N}_{\gamma(z)}\sin (\kappa_1^{\gamma(z)} x)e^{\kappa_2x}|\xi_{\alpha}\rangle,
\label{sx}
\eeqn
with $\kappa_1^{\gamma(z)}=\sqrt{2|m_{\gamma(z)}|/t_x-1/t_x^2}$, $\kappa_2=1/t_x$, and $\mathcal{N}_{\gamma(z)}$ the normalization factors. Spinor $|\xi_{\alpha}\rangle$ statisfies $\tau_z\sigma_ys_x|\xi_{\alpha}\rangle=|\xi_{\alpha}\rangle$. We choose them as
\beqn
&|\xi_1\rangle=\frac{1}{\sqrt{2}}(|+,+,+\rangle+|-,+,-\rangle),\nonumber\\
&|\xi_2\rangle=\frac{1}{\sqrt{2}}(|+,-,-\rangle-|-,-,+\rangle),\nonumber\\
&|\xi_3\rangle=\frac{1}{\sqrt{2}}(|+,-,-\rangle+|-,-,+\rangle),\nonumber\\
&|\xi_4\rangle=\frac{1}{\sqrt{2}}(|-,+,-\rangle-|+,+,+\rangle),
\eeqn
with $|z_1,z_2,z_3\rangle=|\tau_z=z_1\rangle\otimes|\sigma_y=z_2\rangle\otimes|s_x=z_3\rangle$. Projecting $H_p$ into the subspace spanned by these four zero energy states, the surface Hamiltonian can be written as
\beqn
\tilde{H}^{\gamma(z)}=k_y\tilde{\tau}_x\tilde{s}_z\pm k_z\tilde{\tau}_y\tilde{s}_0+\Delta_{x}^{\gamma,z}\tilde{\tau}_z\tilde{s}_0,
\label{sh}
\eeqn
where $\tilde{\tau}$ and $\tilde{s}$ are Pauli matrices acting in zero energy subspace and $\Delta_{x}^{\gamma(z)}=\bar{\Delta}+\Delta_1m_{\gamma(z)}/t_x$ denotes the pairing gap magnitude of the Dirac cones on (100) surface. As a result, the two superconducting gaps $\Delta_{x}^{\gamma}$ and $\Delta_{x}^{z}$ have different magnitudes although
the pairing order parameter $\Delta(\bm k)$ does not depend on $k_z$. This is because surface Dirac electrons at $\Gamma$ and $Z$ points have different decaying coherence length, determined by the bulk electronic bandgap $m_{\gamma(z)}$, which is captured by the pairing potential function $\Delta(x)$ in the process of projection.
Note that the time-reversal symmetry and particle-hole symmetry are intact for surface Hamiltonian $\tilde{H}^{\gamma(z)}$, with given by $\tilde{\mathcal{T}}=i\tilde{s}_y\mathcal{K},\tilde{\mathcal{P}}=\tilde{\tau}_x\mathcal{K}$. Thus, the surface Hamiltonian belongs to DIII symmetry class, which has a $Z_2$ topological classification\cite{Schnyder2008,Ryu2010,Chiu2016}.

Here, although it is not easy to obtain the surface Hamiltonian of the whole Brillouin zone, the topology of the surface BdG Hamiltonian can be fully determined by the sign of the superconducting gap functions $\Delta^{\gamma(z)}_x$ of the Fermi surfaces around $\Gamma$ and $Z$ points. From the phase $\Delta_x^{\gamma}\Delta_x^z>0$ to the phase $\Delta_x^{\gamma}\Delta_x^z<0$, the surface gap must be closed, indicating a surface topological phase transition. When $\Delta_x^{\gamma}\Delta_x^z<0$, (100) surface exists an odd number of Fermi surfaces with negative superconducting pairing. In this case, the (100) surface is expected to be nothing but a TRITSC. Consequently, the topology of (100) surface can be characterized by the surface $Z_2$ topological invariant
\beqn
\label{Z2}
(-1)^{\nu_x}=\sgn({\Delta_{x}^{\gamma}\Delta_{x}^{z}}).
\eeqn
The condition $\nu_x=1$ indicates the realization of surface TRITSC and it leads to 
\beqn
 -2+|m_{z}|/t_x<\Delta_0/\Delta_1<-2+|m_{\gamma}|/t_x.
\eeqn
Thus, the region of topological parameters is proportional to the difference between bulk electronic bandgap $m_{\gamma}$ and $m_{z}$, equaling to $2t_z$.


When taking the open boundary condition of $y$ direction, the low energy Hamiltonian of (010) surface can be obtained through a similar solving process\cite{supp}. This surface Hamiltonian gives rise to the identical physics to that on (100) surface. Analogously, the topology of (010) surface can be characterized by the surface $Z_2$ topological invariant
\beqn
\label{Z2}
(-1)^{\nu_y}=\sgn({\Delta_{y}^{\gamma}\Delta_{y}^{z}}),
\eeqn
with $\Delta_{y}^{\gamma(z)}=\bar{\Delta}+\Delta_1m_{\gamma(z)}/t_y$, which denote the pairing gap magnitude of the two Dirac cones on (010) surface. When $\Delta_{y}^{\gamma}\Delta_{y}^{z}<0$, $\nu_y$ takes value 1, which implies that the (010) surface behaves as a TRITSC.

\begin{figure}
\centering
\includegraphics[width=3.3in]{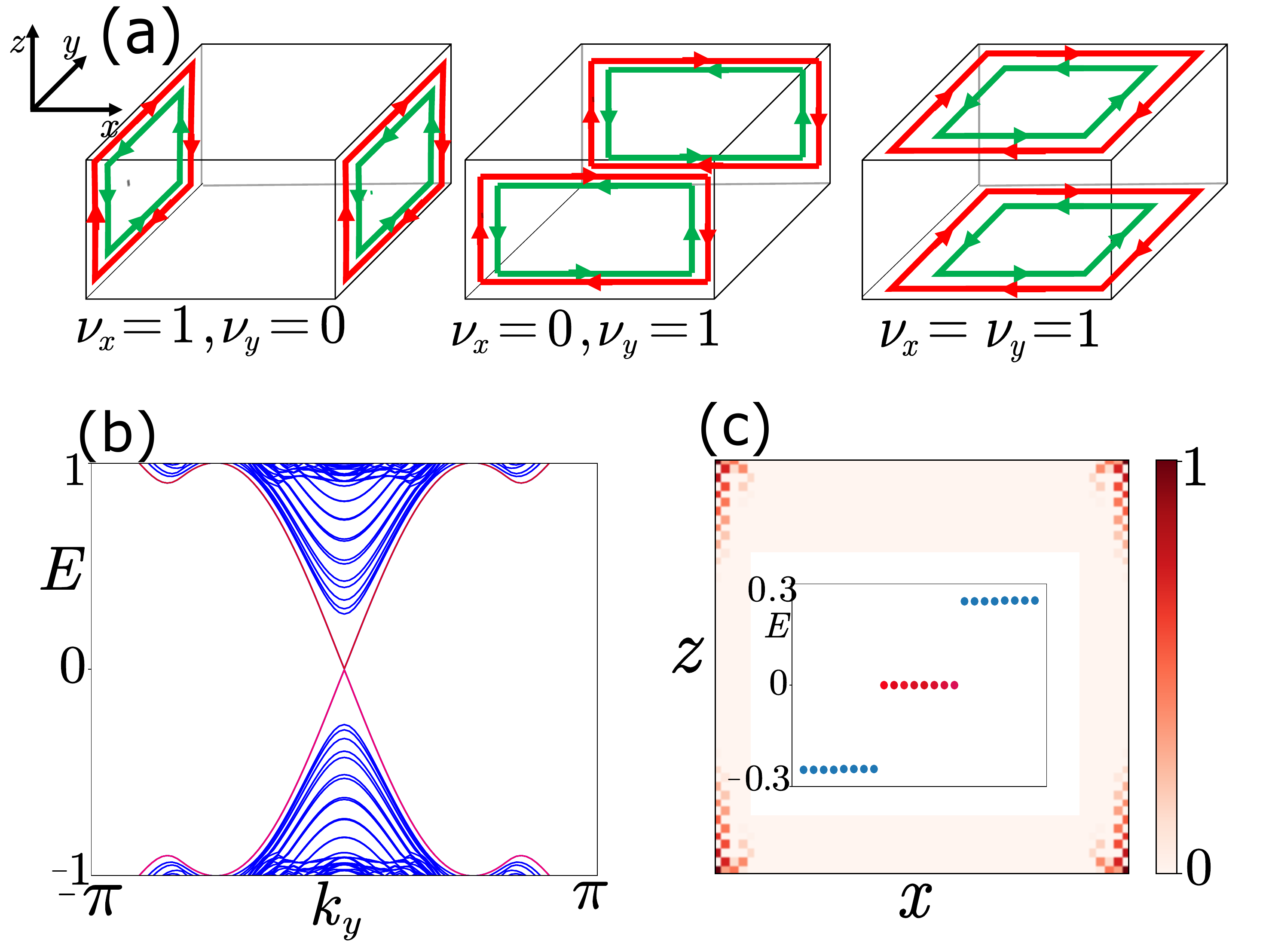}
\caption{(a): The distributions of the HMHMs are schematically plotted under the parameters, yielding different surface $Z_2$ topological invariants $\nu_x$ and $\nu_y$. (b): The energy spectrum of a wire geometry along $y$ is plotted under the parameters $m=2,t_x=t_y=2,t_z=1,\Delta_0=-\Delta_1=-0.5$, with which we have $\nu_x=\nu_y=1$. The in-gap states are of
four-fold degeneracy, which correspond to four pairs of HMHMs. (c): The spatial profile of the MCMs  is plotted when $k_y=0$ in (b) and the inset plots the energies close to zero.}
\label{HS}
\end{figure}


Since 2D TRITSC hosts helical edge Majorana modes and the edge of the surface is the hinge of the 3D system, the surface TRITSC will be manifested by the presence of HMHMs. Therefore, when surface topological invariants $\nu_x=1$ and $\nu_y=0$, there are HMHMs localized at the hinges shared by (100) surface and other surfaces. When surface topological invariants $\nu_x=0$ and $\nu_y=1$, there are HMHMs localized at the hinges shared by (010) surface and other surfaces. Especially, when surface topological invariants $\nu_x=\nu_y=1$, both (100) and (010) surfaces are TRITSCs. The hinges shared by (100) and (010) surfaces have two copies $Z_2$ protected HMHMs, which will couple with each other resulting in no hinge modes.
Thus, in this case, there are HMHMs localized at the hinges shared by (001) surface and other surfaces. In Fig.~\ref{HS}(a), we provide the schematic plot of the HMHMs for these three different cases. In Fig.~\ref{HS}(b)(c), we verify the existence of the HMHMs through numerical calculation for the case $\nu_x=\nu_y=1$. Naturally,
the whole system is a second-order TSC in DIII symmetry class due to the existence of HMHMs.

The HMHMs can also be characterized by the winding of the Wilson loop spectrum for a slab geometry\cite{Zhang2019}. For instance, the HMHMs along $k_y$  in Fig.~\ref{HS}(b) can be characterized by the winding of the Wannier spectrum along $k_y$, obtained by performing Wilson loop along $k_x$  for a slab geometry on the (001) surface\cite{supp}. This winding of the Wannier spectrum unambiguously signals the existence of HMHMs.

\subsection{Second-order TSC in D symmetry class}
Up to now, the time-reversal symmetry that protects the HMHMs is preserved. When further applying an in-plane magnetic field to break time-reversal symmetry, another interesting surface topological phase transition can occur. Without loss of generality, considering the Zeeman term $V_x\tau_zs_x$ induced by the magnetic field along $x$ direction, then the degenerate surface states on (100) surface have Zeeman splitting. Similar to strong TIs, this Zeeman term produces no influence for the surface states on (010) surface\cite{Khalaf2018,Wu2020a}. Consequently, the superconducting pairing gaps $\Delta_{x}^{\gamma(z)}$ are split to two-gap amplitudes $|\Delta_{x}^{\gamma(z)}\pm V_x|$. Obviously, increasing $V_x$ from zero, the surface gap first decreases to zero at $\{|\Delta_{x}^{\gamma}|, |\Delta_{x}^{z}|\}_{\text{min}}= |V_x|$ and then reopens with further increasing $V_x$. This indicates another surface topological phase transition, which is associated with the gap closing and reopening of the non-degenerate surface Dirac cone. It is known that the Chern number will be changed one when the energy gap of the massive Dirac cone is closed and reopened \cite{Bernevig2013}. On the other hand, no matter the topological invariant $\nu_x=1$ or $\nu_x=0$ in the absence of the magnetic field, the  surface Chern number for whole surface bands is zero due to time-reversal symmetry. Thus, in the presence of finite magnetic field, the (100) surface can be driven to a surface chiral TSC characterized by nontrivial surface Chern number.



\begin{figure}
\centering
\includegraphics[width=3.3in]{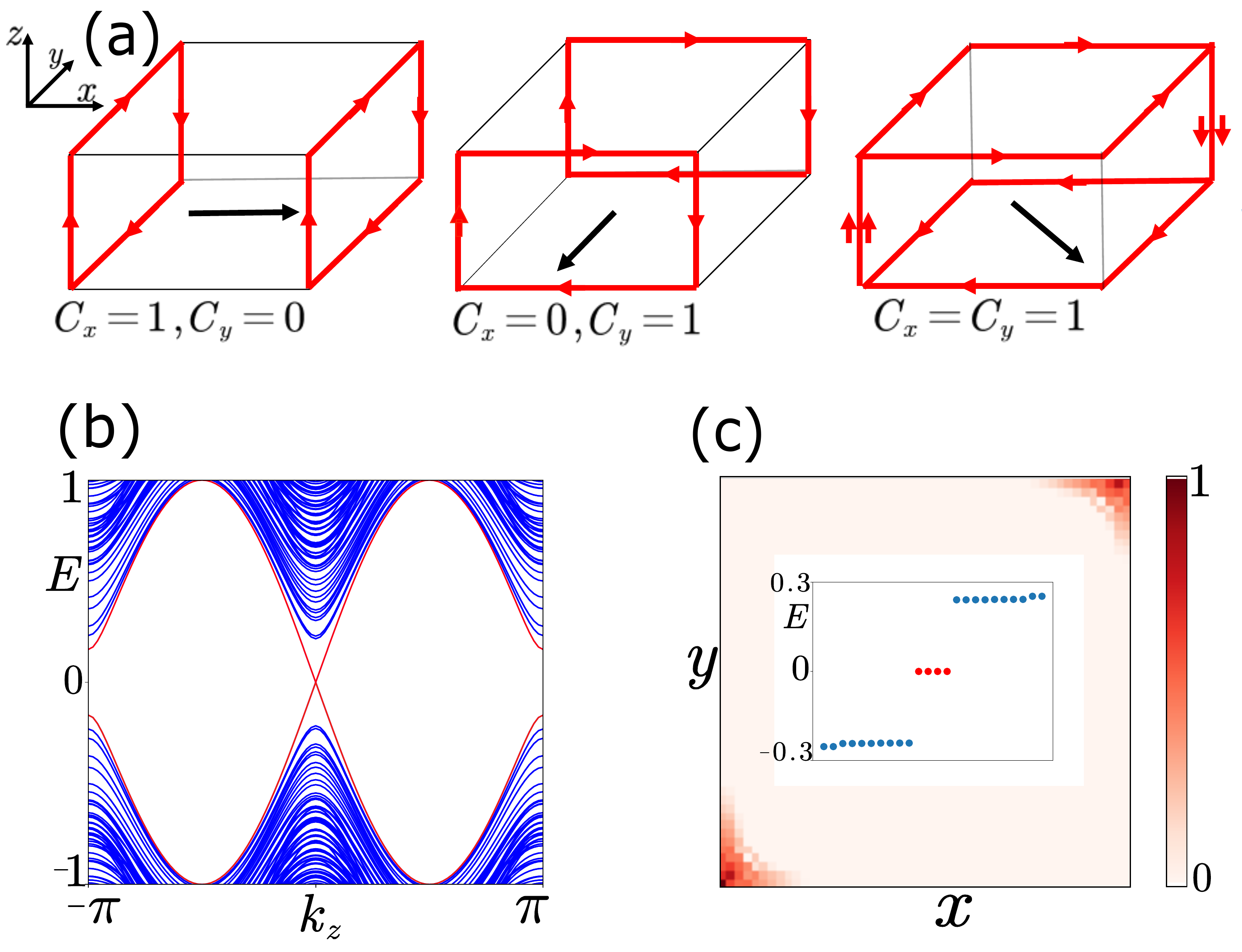}
\caption{(a): The distributions of the CMHMs are schematically plotted under the parameters, yieliding different topological invariants $C_x$ and $C_y$. The black arrow indicates the direction of in-plane magnetic field. It is noted that the CMHMs on opposite surface have identical chirality.  (b): The energy spectrum of a wire geometry along $z$ is plotted under the parameters $m=2,t_x=t_y=2,t_z=1,\Delta_0=-0.1, \Delta_1=0.4,V_x=-V_y=0.3$, which correspond to the topological invariants $\nu_x=\nu_y=0, C_x=C_y=1$. (c): The spatial profile of the MCMs when $k_z=0$ in (b) is plotted and the inset plots the energies close to zero.}
\label{CS}
\end{figure}

To vertify above physical pircture, we caclulate surface Chern number $C_x$ from  low energy surface Hamiltonian. Directly, the influence of the Zeeman term $V_x\tau_zs_x$ for surface states can be taken into account by projecting it into the subspace spanned by zero energy states in Eq.~\eqref{sx}. Performing this projection calculation, we obtain additional term $V_x\tilde{\tau}_z\tilde{s}_z$. Thus, the surface Hamiltonian $\tilde{H}^{\gamma(z)}$ in Eq.~\eqref{sh} transforms into the form
\beqn
\tilde{H}^{\gamma(z)}=k_y\tilde{\tau}_x\tilde{s}_z\pm k_z\tilde{\tau}_y\tilde{s}_0+\Delta_{x}^{\gamma(z)}\tilde{\tau}_z\tilde{s}_0+V_x\tilde{\tau}_z\tilde{s}_z.
\eeqn
Obviously, the surface Hamiltonian is block-diagonal in $\tilde{s}$ space and each block describes a massive Dirac cone, which has been well known to contribute half Chern number, depending on the sign of Dirac mass\cite{Qi2008,Bernevig2013}.
As a result, the Chern number for the whole surface bands is $C_x=C_x^{\gamma}+C_x^{z}$, in which
\beqn
&C_{x}^{\gamma}=(\text{sgn}[\Delta_x^{\gamma}+V_x]-\text{sgn}[\Delta_x^{\gamma}-V_x])/2, \nonumber\\
&C_{x}^{z}=(\text{sgn}[\Delta_x^z-V_x]-\text{sgn}[\Delta_x^z+V_x])/2.
\eeqn
It can be verified that under the condition $\{|\Delta_{x}^{\gamma}|, |\Delta_{x}^{z}|\}_{\text{min}}<V_x<\{|\Delta_{x}^{\gamma}|, |\Delta_{x}^{z}|\}_{\text{max}}$, surface Chern number $C_x$ is  1 or -1 and it can be expressed as
\beqn
C_x=\text{sgn}[V_x]\text{sgn}[|\Delta_x^{z}|-|\Delta_x^{\gamma}|].
\label{chx}
\eeqn
Otherwise, it takes the value 0. Thus, the (100) surface behaves as a chiral TSC under appropriate parameters condition. Similarly, the surface chiral TSC will be manifested by the presence of CMHMs for the 3D system.



When applying magnetic field along $y$ direction with magnitude $V_y$, (010) surface has identical physics as that on (100) surface. Under the parameters condition
$\{|\Delta_{y}^{g}|, |\Delta_{y}^{z}|\}_{\text{min}}<V_y<\{|\Delta_{y}^{g}|, |\Delta_{y}^{z}|\}_{\text{max}}$, the surface Chern number $C_y$ can be written as\cite{supp}
\beqn
C_y=-\text{sgn}[V_y]\text{sgn}[|\Delta_y^{z}|-|\Delta_y^{\gamma}|].
\label{chy}
\eeqn
As we can see in Eq.~\eqref{chx} and Eq.~\eqref{chy}, the chirality of the surface chiral TSC can be tuned by reversing the direction of the magnetic field.

Consequently, the side faces can be independently tuned to a surface TSC in DIII or D symmetry class by considering the interplay between magnetic fields and superconductors. These surface TSCs are characterized by surface topological invariants $\{\nu_x,\nu_y,C_x,C_y\}$. In the absence of magnetic fields, $C_x=C_y=0$ and only HMHMs can appear at certain hinges. In the presence of magnetic fields and under the condition $\nu_x=\nu_y=0$,  only CMHMs can appear at certain hinges. In this case, we provide the schematic plot of the CMHMs characterized by topological invariants $(C_x,C_y)$ in Fig.~\ref{CS}(a). Particularly, when $C_x=C_y=1$, both (100) and (010) surfaces are chiral TSC. There are two copies CMHMs at the hinges shared by adjacent side faces. These CMHMs will couple with each other when they have opposite chirality. On the contrary, the CMHMs with the same chirality will survive, resulting in two-channels CMHMs at certain hinges. This case can be verified by the numerical calculations, as shown in Fig.~\ref{CS}(b)(c). Remarkably,  the HMHMs and CMHMs can coexist at different hinges when the adjacent side faces behave as TRITSC and chiral TSC,  respectively \cite{supp}.


\section{Third-order TSC}

Similar to that second-order TSCs can be realized by performing surface TSCs, a third-order TSC with MCMs can emerge from a 1D hinge TSC. In the following, we theoretically design a model of third-order TSC in D symmetry class to perform this visualized principle and provide topological characterization for this higher-order phase.

\begin{figure}
\centering
\includegraphics[width=3.2in]{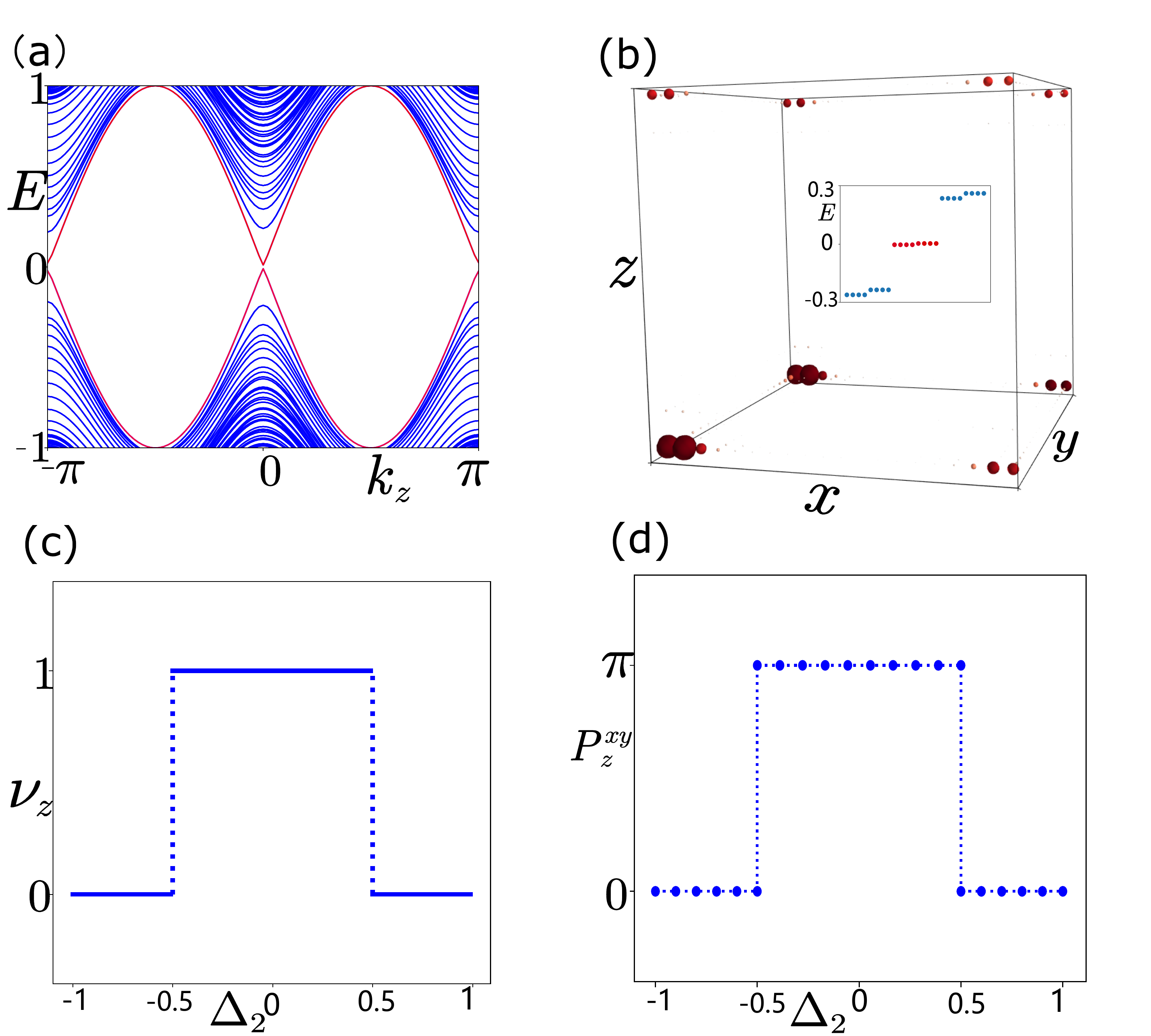}
\caption{(a): The enegy spectrum of a wire geometry along $z$ under the parameters $m=2, t_x=2.5,t_y=1,t_z=0.3,\Delta_0=-\Delta_1=-0.5$. (b): The spatial distribution of the eight MCMs, which are obtained by considering the mass term $0.5\cos k_z\tau_xs_y$ to gap the HMHMs in (a). The inset shows energies close to zero. (c): The topological invariant $\nu_z$ is the function of $\Delta_2$. (d): The numerical calculation of the nested polarization $P_z^{xy}$, as a function of $\Delta_2$.}
\label{pf}
\end{figure}

Our model of third-order TSC is realized based on the novel phase for $H_{\rm BdG}(\bm k)$ with model parameters, in which $\nu_x=\nu_y=0$, but $\Delta_{x}^{\gamma}\Delta_{y}^{\gamma}<0$ and $\Delta_{x}^z\Delta_{y}^z<0$.
In this case, although both (100) and (010) surfaces are topologically trivial, the sign reversal of the pairing gaps between  adjacent side faces will lead to HMHMs, which can be understood from the topological domain walls problem\cite{Teo2010}.   It is noted that because the pairing gaps of Dirac cones  have sign reversal at $k_z=0$ and $k_z=\pi$ simultaneously, the surface domain walls bind the hinges shared by (100) and (010) surfaces with two pairs of HMHMs.
As shown in Fig.~\ref{pf}(a), two pairs of HMHMs cross at $k_z=0$ and $k_z=\pi$, respectively.

Based on this novel phase, we theoretically consider the pairing term $\tilde{\Delta}(k_z)=(\Delta_2+\Delta_3\cos k_z)\tau_xs_y$, breaking time-reversal symmetry, to gap the two pairs of HMHMs. The hinge gap can be obtained by projecting $\tilde{\Delta}(k_z)$ into the subspace of the HMHMs. Notably, at $k_z=0,\pi$, this pairing term is momentum-independent. Thus, the HMHMs crossing at $k_z=0$ and $k_z=\pi$ are gaped with the magnitude $\Delta_2+\Delta_3$ and $\Delta_2-\Delta_3$, respectively.
Now, we study the topology of the hinges shared by (100) and (010) surfaces. These hinges belong to D symmetry class and have a $Z_2$ topological classification\cite{Schnyder2008,Ryu2010,Chiu2016}.
The topological invariant is explicitly expressed as\cite{Kitaev2001}
\beqn
(-1)^{\nu_z}=\sgn[\frac{\text{Pf}(H_{\text{hinge}}( k_z=0))}{\text{Pf}(H_{\text{hinge}}( k_z=\pi))} ],
\eeqn
where $\text{Pf }$ is the Pfaffian of a antisymmetric matrix and $H_{\text{hinge}}$ denotes the hinge Hamiltonian written in Majorana basis. The $Z_2$ topological invariant $\nu_z$ is 1 for the nontrivial phase and 0 for the trivial phase. At $ k_z=0,\pi$, the sign of Pfaffian is completely determined by the sign of energy gap\cite{Kitaev2001,Ghosh2010}. Consequently, the definition of $\nu_z$ transforms into the form: $(-1)^{\nu_z}=\sgn[(\Delta_2+\Delta_3)(\Delta_2-\Delta_3)]$, which takes the value 0 when $|\Delta_{2}/\Delta_{3}|>1$ and the value 1 when $|\Delta_{2}/\Delta_{3}|<1$. From the phase $|\Delta_2|>|\Delta_3|$ to the phase $|\Delta_2|<|\Delta_3|$, the hinge energy gap is closed, indicating a hinge topological phase transition. Thus, the hinges are 1D TSC in D symmetry class under appropriate parameters condition. Correspondingly, these 1D hinge TSCs will be manifested by the presence of MCMs when taking open boundary conditions of all directions. As shown in Fig.~\ref{pf}(b), there are one MCM located at each corner in real space. The presence of MCMs shows that the whole system is a third-order TSC.

Besides the Pfaffian formula characterization, the hinge topology can also be revealed by the well-known nested Wilson loop topological invariants based on the equivalent topology between Wannier band and boundaries\cite{Benalcazar2017,Benalcazar2017a}. Performing Wilson loop along $k_x$ and then performing nested Wilson loop along $k_y,k_z$ in sequence, we obtain the nested Wilson loop polarization
$P_z^{xy}$\cite{Benalcazar2017,Benalcazar2017a,supp}. This topological index is quantized to $\pi$ for the third-order TSC phase and 0 for the trivial phase. The numerical result is shown in Fig.~\ref{pf}(d), which exhibits a sharp topological phase transition agreeing with the Pfaffian characterization shown in Fig.~\ref{pf}(c). Therefore, the nested polarization $P_z^{xy}$ can also characterize this third-order TSC as a $Z_2$ topological invariant.

\section{Discussion and conclusion}
For simplicity, we have taken the chemical potential to be zero in our discussion. However, our conclusions do not depend on this condition. The Majorana hinge modes and MCMs can survive in the presence of finite
chemical potential as long as the surface and hinge energy gap preserve, respectively. Similarly, these boundary states are robust against the perturbations preserving boundary gap, even they couple the two Dirac cones on the same side face.

In conclusion, based on the weak TIs with band inversion at both $\Gamma$ and Z points, we consider the interplay between superconductors and magnetic fields in this system to realize second-order TSCs featuring HMHMs or CMHMs and third-order TSC featuring MCMs. Essentially, the chiral Majorana hinge modes and MCMs in our models origin from the surface and hinge first-order TSC, respectively.
Correspondingly, these higher-order TSCs can be characterized by corresponding boundary topological invariants.

\section*{Acknowledgement}
We would like to thank Chao-Xing Liu for useful discussions. This work is supported by NSFC (Grant No.11674114), NSFC (Grant No.12074133) and National Key R\&D Program of China (Grant No. 2016YFA0401003).

\appendix
\begin{widetext}

\section{The topology of (010) surface}
In this section, we focus on the topology of (010) surface. To obtain the surface Hamiltonian, we take the open boundary condition of $y$ direction. Under this condition,
we replace $k_y\rightarrow -i\partial_y$, then the BDG Hamiltonian expanding at $\Gamma$ and $Z$ points can be decomposed as $H_{\rm BdG}^{\gamma(z)}=H_0^{\gamma,z}(y)+H_p^{\gamma(z)}(y,k_x,k_z)$, in which
\beqn
H_0^{\gamma(z)}(y)=(m_{\gamma(z)}-\frac{t_y}{2}\partial_y^2)\tau_{z}\sigma_z-i\partial_y\tau_z\sigma_x s_y,\nonumber\\
H_p^{\gamma(z)}(y,k_x,k_z)=k_x\sigma_x s_x\pm k_z\sigma_x s_z+\Delta(y)\tau_y s_y,
\eeqn
where $\Delta(y)$ is defined as $\bar{\Delta}+\frac{\Delta_1\partial_y^2}{2}$ and the insignificant terms $k_x^2$ and $k_z^2$ involved have been omitted. Now we solve the zero energy states of $H_0^{\gamma(z)}$ and consider $H_p^{\gamma(z)}$ as a perturbation to extract the surface Hamiltonian.

Solving the eigen equation $H_0^{\gamma(z)}|\psi_{\alpha}^{\gamma(z)}(y)\rangle=0$ under the boundary conditions $|\psi_{\alpha}^{\gamma(z)}(y=0)\rangle=|\psi_{\alpha}^{\gamma(z)}(y=-\infty)\rangle=0$, there are four zero energy states. The wave function of these zero states can be written as
\beqn
|\psi_{\alpha}^{\gamma(z)}(y)\rangle=\mathcal{N}_{\gamma,z}\sin (\kappa_1^{\gamma(z)} y)e^{\kappa_2y}|\xi_{\alpha}\rangle,
\label{zew}
\eeqn
with $\kappa_1^{\gamma(z)}=\sqrt{2|m_{\gamma(z)}|/t_y-1/t_y^2}$, $\kappa_2=1/t_y$,  and $\mathcal{N}_{\gamma(z)}$ the normalization factors. Spinor $|\xi_{\alpha}\rangle$ statisfies $\sigma_ys_y|\xi_{\alpha}\rangle=|\xi_{\alpha}\rangle$. We choose them as
\beqn
&|\xi_1\rangle=|\tau_y=+1\rangle\otimes|\sigma_y=+1\rangle\otimes|s_y=+1\rangle,\nonumber\\
&|\xi_2\rangle=|\tau_y=-1\rangle\otimes|\sigma_y=-1\rangle\otimes|s_y=-1\rangle,\nonumber\\
&|\xi_3\rangle=|\tau_y=+1\rangle\otimes|\sigma_y=-1\rangle\otimes|s_y=-1\rangle,\nonumber\\
&|\xi_4\rangle=|\tau_y=-1\rangle\otimes|\sigma_y=+1\rangle\otimes|s_y=+1\rangle.
\eeqn
Projecting $H_p$ into the subspace spanned by these four zero energy states, the low energy surface Hamiltonian  can be written as
\beqn
\tilde{H}^{\gamma(z)}= -k_x\tilde{\tau}_x\tilde{s}_0\pm k_z\tilde{\tau}_y\tilde{s}_z+\Delta_{y}^{\gamma,z}\tilde{\tau}_z\tilde{s}_0,
\label{hy}
\eeqn
where $\Delta_{y}^{\gamma(z)}=\bar{\Delta}+2\Delta_1m_{\gamma(z)}/t_y$ denote the superconducting pairing gap of the Dirac cones, Pauli matrices $\tilde{\tau}$ and $\tilde{s}$ act in the zero energy states space. When $\Delta_y^{\gamma}\Delta_y^{z}<0$, (010) surface exists an odd number of Fermi surfaces with negative superconducting pairing. In this case, the (010) surface behaves as a TRITSC. As a result, the topology of (010) surface can be characterized by the surface $Z_2$ topological invariant
\beqn
\label{Z2}
(-1)^{\nu_y}=\sgn({\Delta_{y}^{\gamma}\Delta_{y}^{z}}).
\eeqn
The condition $\nu_y=1$ indicates  the realization of surface TRITSC and it leads to 
\beqn
 -2+|m_{z}|/t_y<\Delta_0/\Delta_1<-2+|m_{\gamma}|/t_y.
\eeqn

Further considering the Zeeman term $V_ys_y$ induced by the magnetic field along y direction, then the surface Hamiltonian in Eq.~\eqref{hy} transforms into the form
\beqn
\tilde{H}^{\gamma(z)}=-k_x\tilde{\tau}_x\tilde{s}_0\pm k_z\tilde{\tau}_y\tilde{s}_z+\Delta_{y}^{\gamma(z)}\tilde{\tau}_z\tilde{s}_0+V_y\tilde{\tau}_z\tilde{s}_z.
\eeqn
Obviously, the obtained surface Hamiltonian is block-diagonal and each block contributes half Chern number. Therefore, the Chern number for the whole surface bands is $C_y=C_y^{\gamma}+C_y^{z}$, in which
\beqn
&C_{y}^{\gamma}=(\text{sgn}[\Delta_y^{\gamma}-V_y]-\text{sgn}[\Delta_y^{\gamma}+V_y])/2, \nonumber\\
&C_{y}^{z}=(\text{sgn}[\Delta_y^{z}+V_y]-\text{sgn}[\Delta_y^{z}-V_y])/2.
\eeqn
Under the condition $\{|\Delta_{y}^{\gamma}|, |\Delta_{y}^{z}|\}_{\text{min}}<V_y<\{|\Delta_{y}^{\gamma}|, |\Delta_{y}^{z}|\}_{\text{max}}$, $C_y$ takes value $1$ or $-1$ and it can be expressed as
\beqn
C_y=-\text{sgn}[V_y]\text{sgn}[|\Delta_y^{\gamma}|-|\Delta_y^{z}|].
\label{ch}
\eeqn
Otherwise, it takes the value 0.

\begin{figure}
\centering
\includegraphics[width=6.2in]{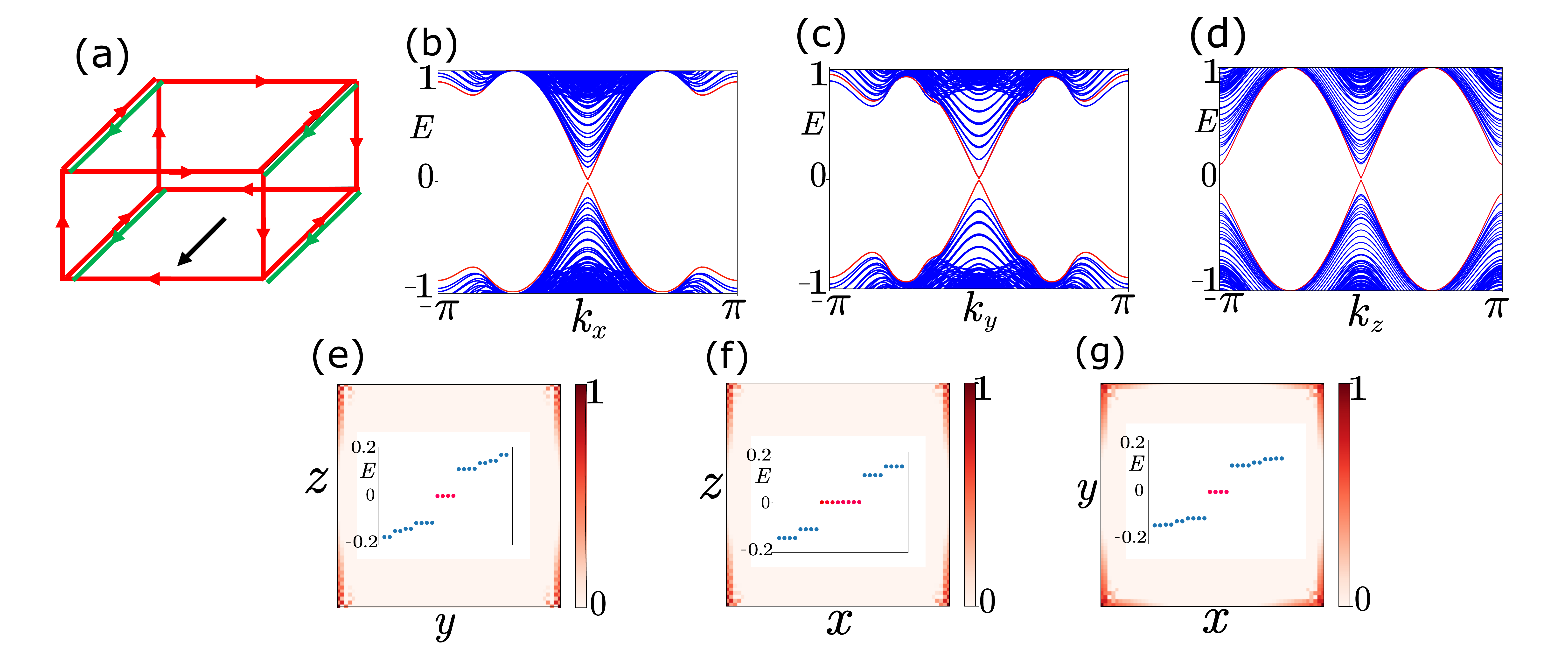}
\caption{Common parameters are taken as $m=2,t_x=t_y=2,t_z=1,\Delta_0=-0.3, \Delta_1=0.4,V_y=0.2$. These parameters
yield surface topological invariants $\nu_x=1,\nu_y=0,C_x=0,C_y=1$. (a): Schematic plot of the distribution of CMHMs and HMHMs. (b-d): The energy spectrum of a wire geometry along $x,y,z$ directions, respectively. (e-g): The spatial profile of the MCMs when $k_x=0,k_y=0,k_z=0$ in (b-d) and the insets plot the energies close to zero.}
\label{CHS}
\end{figure}

\section{The coexistence of HMHMs and CMHMs }
In this section, we show that the HMHMs and CHMHMs can coexist in our system. This scenario can be achieved when the adjacent side faces behave as TRITSC and chiral TSC, respectively. For example, applying finite
magnetic filed along y direction to the second-order TSC characterized by topological invariants $\nu_x=1,\nu_y=0$, then the (010) surface is driven to a chiral TSC. In this case, we obtain second-order TSC characterized by topological invariants $\nu_x=1,\nu_y=0,C_x=0,C_y=1$. As schematically plotted in Fig.~\ref{CHS}(a), HMHMs and CMHMs can coexist in this system. In Fig.~\ref{CHS}(b-g), we check the distribution of the HMHMs and CMHMs in (a) through numerical calculations.

\begin{figure}
\centering
\includegraphics[width=6.2in]{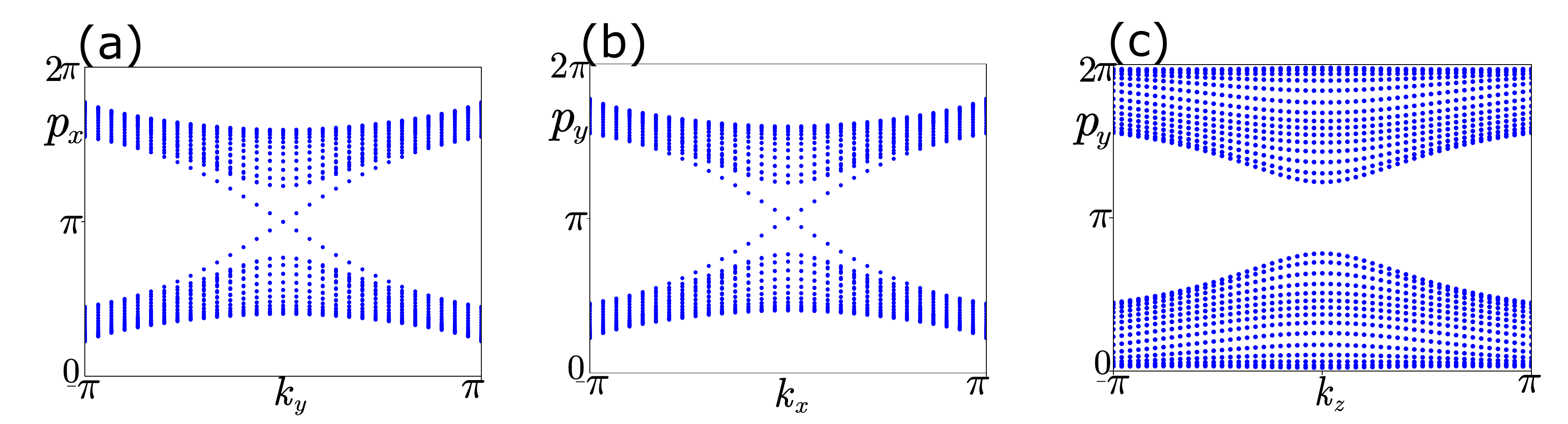}
\caption{Common parameters are taken as $m=2,t_x=t_y=2,t_z=1,\Delta_0=-0.5, \Delta_1=0.5$. These parameters yield surface topological invariants $\nu_x=\nu_y=1$. (a): The Wannier spectrum $p_x(k_y)$ is plotted by performing Wilson loop along $k_x$ for the slab geometry on (001) surface. (b): The Wannier spectrum $p_y(k_x)$ is plotted by performing Wilson loop along $k_y$ for the slab geometry on (001) surface. (c): The Wannier spectrum $p_y(k_z)$ is plotted by performing Wilson loop along $k_y$ for the slab geometry on (100) surface.}
\label{WL}
\end{figure}

\section{Wilson loop and nested Wilson loop}
In this section, we introduce the standard procedure of calculating Wilson loop and nested Wilson loop, which can characterize the HMHMs and MCMs, respectively. Considering Wilson loop along $k_x$, the Wannier center can be obtained by diagonalizing the Wilson loop matrix
\beqn
W_{x,\bm{k}}|\nu_{x,\bm k}^{j}\rangle=\exp[i p_{x}^{j}(k_y,k_z) ]|\nu_{x,\bm k}^{j}\rangle.
\label{wx}
\eeqn
Here, $j\in\{1,\ldots,N_\text{occ}\}$ with $N_\text{occ}$ the number of occupied states, $W_{x,\bm{k}}$ is $N_\text{occ}\times N_\text{occ}$ matrix with matrix element $W_{x,\bm{k}}^{nm}= \lim_{N_x \rightarrow \infty} \, \bra{n, \bm{k}+\bm{G}_x} \Big[ \prod_{s=1}^{N_x-1} P_x(\bm k_s) \Big] \ket{m, \bm{k}}$, where
$\ket{m, \bm{k}}$ is the m-th occupied state at $\bm k$ and $\bm{G}_x$ denotes reciprocal vector along $x$ direction, the projector operator $P_x(\bm k_s)$ for the occupied states is defined as $P_x(\bm k_s)=\sum_{j=1}^{N_{ occ}} \ket{j, \bm{k})_s} \bra{j, \bm{k}_s}$, with $\bm k_s=\bm{k}+\frac{s}{N_x} \bm{G}_x$.

The HMHMs can be characterized by the winding of the Wannier center obtained for the system with a slab geometry. For example, the second-order TSC phase characterized by topological invariants $\nu_x=\nu_y=1$ hosts HMHMs at the hinges shared by (001) surface and side surfaces. To characterize these HMHMs, we consider the slab geometry on (001) surface with the open boundary condition of $z$ direction. For this system, the Wannier center $p_{x}(k_y)$ and $p_{y}(k_x)$ are obtained by performing Wilson loop along $k_x$ and $k_y$, respectively. As shown in Fig.~\ref{WL}(a)(b), the Wannier center $p_{x}(k_y)$ and $p_{y}(k_x)$ exhibit a helical winding as the HMHMs, which signal the existence of the HMHMs. However, when considering the slab geometry on (100) surface and perform Wilson loop along $k_y$, the Wannier spectrum $p_y(k_z)$ is gapped, as shown in Fig.~\ref{WL}(c).

\begin{figure}
\centering
\includegraphics[width=4.2in]{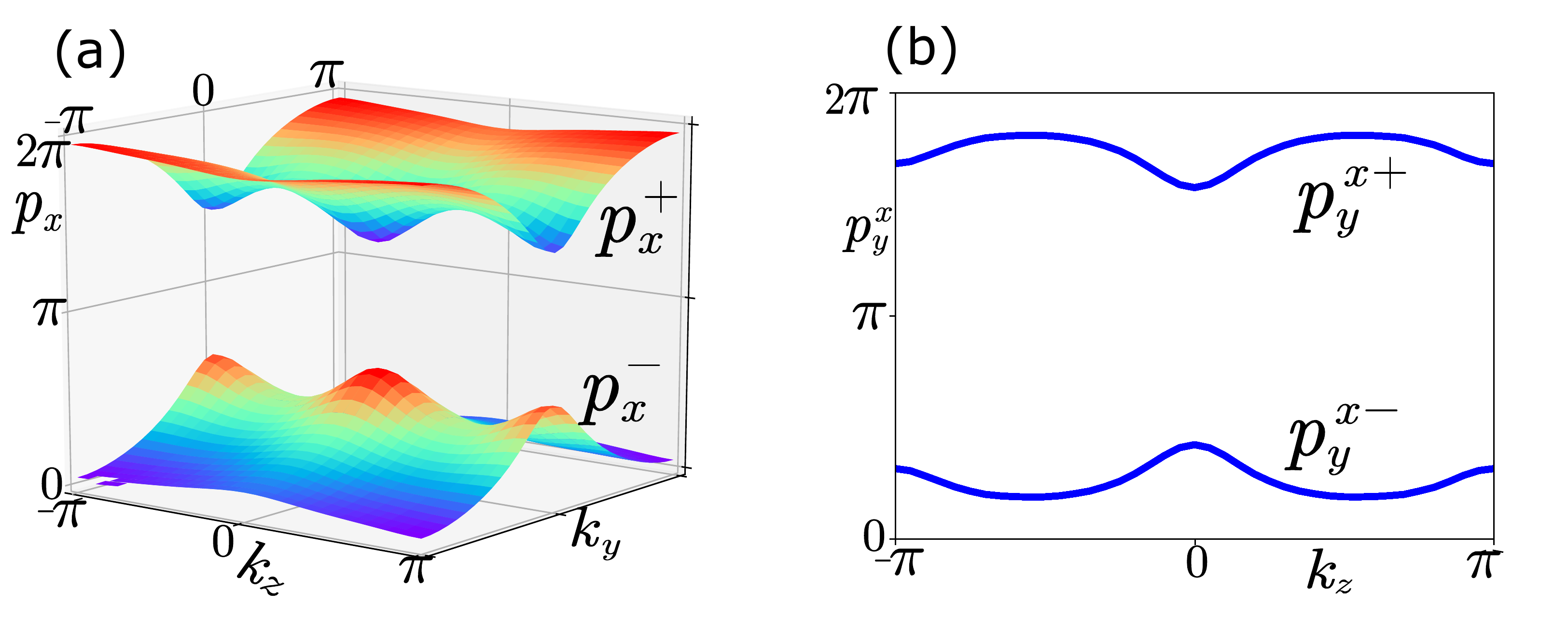}
\caption{The Wannier bands are obtained by calculating Wilson loop and nested Wilson loop based on the  Hamiltonian $H_{\text{BDG}}(\bm k)+\bar{\Delta}(k_z)$. The parameters are taken as $m=2,t_x=2.5,t_y=1,t_z=0.3,\Delta_0=-\Delta_1=-\Delta_3=-0.5,\Delta_2=0$. (a): 2D Waniier spectrum $p_x(k_y,k_z)$ is plotted by performing Wilson loop along $k_x$. (b):Nested Wannier spectrum $p_y^x(k_z)$ is plotted by performing nested Wilson loop along $k_y$.}
\label{nwl}
\end{figure}

To characterize the third-order TSC in main text, we perform Wilson loop along $k_x$ and nested Wilson loop along $k_y,k_z$ in sequence for eight bands Hamiltoian $H_{\text{BDG}}(\bm k)+\bar{\Delta}(k_z)$ in momentum space. Performing Wilson loop along $k_x$ as Eq.~\eqref{wx}, we obtain 2D Wannier band $p_{x}(k_y,k_z)$, which is gapped when the energy spectrum of (100) surface is gapped. As shown in Fig.~\ref{nwl}(a), the Wannier gap splits the Wannier spectrum $p_{x}(k_y,k_z)$ into two sectors $p_x^{\pm}(k_y,k_z)$ associated with corresponding eigenstates $|\nu_{x,\bm k}^{\pm,r}\rangle$, with $r=1,2$. Based on the eigenstates $|\nu_{x,\bm k}^{+,r}\rangle$, we construct the nested Wannier basis
\beqn
|w_{x,\bm{k}}^{+,r} \rangle= \sum_{j=1}^{4}|j,\bm k\rangle [\nu_{x,\bm{k}}^{+,r}]^j,
\eeqn
where $[\nu_{x,\bm{k}}^{+,r}]^j$ is the $j$ component of the eigenstate $|\nu_{x,\bm{k}}^{+,r}\rangle$.  Calculating the nested Wilson loop along $k_y$, the nested Wannier spectrum can be obtained by diagonalizing the matrix
\beqn
\mathcal{W}_{y,\bm{k}}^{+}|\eta_{y,\bm k}^{+}\rangle=\exp[i p_{y}^{x}(k_z) ]|\eta_{y,\bm k}^{+}\rangle.
\eeqn
Here, $\mathcal{W}_{y,\bm{k}}^{+}$ is $2\times 2$ matrix with matrix element $(\mathcal{W}_{y,\bm{k}}^{+})^{rr^{'}}= \lim_{N_y \rightarrow \infty} \langle w_{x,\bm{k}+\bm{G}_{y}}^{+,r} \Big[ \prod_{q=1}^{N_y-1} P(\bm{k}_q) \Big] \ket{w_{x,\bm{k}}^{+,r^{'}}}$, where $\bm{G}_y$ denotes reciprocal vector along $y$ direction, the projecting operator $P(\bm{k}_q)$ is defined as $P(\bm{k}_q)=\sum_{r=1}^{2} |w_{x,\bm{k}}^{+,r}\rangle \langle w_{x,\bm{k}}^{+,r} |$, with $\bm{k}_q=\bm{k}+\frac{q}{N_y} \bm{G}_y$. The nested Wannier band $p_{y}^{x}(k_z)$ is gapped when the energy spectrum of the hinges shared by (100) and (010) surfaces is gapped. As shown in Fig.~\ref{nwl}(b), the Wannier gap splits the nested Wannier spectrum $p_{y}^{x}(k_z)$ into two sectors $p_{y}^{x\pm}$ with corresponding eigenstates $|\eta_{y,\bm k}^{+,\pm}\rangle$. We choose Wannier sector $p_{y}^{x+}$ to construct Wannier basis
\beqn
|\tilde{w}_{y,\bm{k}}^{+,+} \rangle= \sum_{r=1}^{2}|w_{x,\bm{k}}^{+,r}\rangle [\eta_{y,\bm k}^{+,+}]^r,
\eeqn
where $[\eta_{j,\bm k}^{+,+}]^r$ is the $r$ component of the eigenstate $|\eta_{y,\bm k}^{+,+}\rangle$. The nested Wilson loop along $k_z$ can be performed by diagonalizing the matrix
\beqn
\tilde{\mathcal{W}}_{z,\bm{k}}^{+,+}|\tilde{\eta}_{z,\bm k}^{+,+}\rangle=\exp[i p_{z}^{xy} ]|\tilde{\eta}_{z,\bm k}^{+,+}\rangle.
\eeqn
Here, $\tilde{\mathcal{W}}_{z,\bm{k}}^{+,+}=\lim_{N_z \rightarrow \infty} \langle\tilde{w}_{y,\bm{k}+\bm G_{z}}^{+,+} \Big[ \prod_{l=1}^{N_z-1} P_z(\bm{k}_l) \Big] |\tilde{w}_{y,\bm k}^{+,+}\rangle$, where $\bm{G}_z$ denotes reciprocal vector along $z$ direction, the projecting operator $P_z(\bm{k}_l)$ is defined as $P_z(\bm{k}_l)= \ket{\tilde{\eta}_{z,\bm k}^{+,+}} \langle\tilde{\eta}_{z,\bm k}^{+,+} |$, with $\bm{k}_l=\bm{k}+\frac{l}{N_z} \bm{G}_z$. Finally, we obtatin nested polarization
\beqn
P_{z}^{xy}=-i\frac{1}{N_xN_y}\sum_{k_x,k_y}\text{log}[p_{z}^{xy}].
\eeqn
 The numerical result shows that $ P_{z}^{xy}$ is exactly quantized to $\pi$ for the 3th-TSC phase. Here, it is noted that the order of the nested Wilson loops $W_{x}\rightarrow \mathcal{W}_{y}$ is arbitrary to characterize the 3th-TSC.

\end{widetext}

\end{document}